# Phase-Aware Speech Enhancement with a Recurrent Two Stage Network

*Juntae Kim*[1] *and Jaesung Bae*[2]

[1]kakao, Gyeonggi-do, Korea
[2]NCSOFT, Gyeonggi-do, Korea
`jtkim@kaist.ac.kr, jaesungbae@ncsoft.com`

## Abstract

We propose a neural network-based speech enhancement (SE) method called the phase-aware recurrent two stage network (rTSN). The rTSN is an extension of our previously proposed two stage network (TSN) framework. This TSN framework was equipped with a boosting strategy (BS) that initially estimates the multiple base predictions (MBPs) from a prior neural network (pri-NN) and then the MBPs are aggregated by a posterior neural network (post-NN) to obtain the final prediction. The TSN outperformed various state-of-the-art methods; however, it adopted the simple deep neural network as pri-NN. We have found that the pri-NN affects the performance (in perceptual quality), more than post-NN; therefore we adopted the long short-term memory recurrent neural network (LSTM-RNN) as pri-NN to boost the context information usage within speech signals. Further, the TSN framework did not consider the phase reconstruction, though phase information affected the perceptual quality. Therefore, we proposed to adopt the phase reconstruction method based on the Griffin-Lim algorithm. Finally, we evaluated rTSN with baselines such as TSN in perceptual quality related metrics as well as the phone recognition error rate.

## 1. Introduction

The objective of speech enhancement (SE) is the separation of clean speech signals from the noisy and corrupted speech signals. SE is widely used as a front-end system for various speech-related applications such as hearing aids and speech recognition [1]. Recently, SE with deep learning framework had outperformed conventional methods [2-4], specifically in non-stationary environments. The basic SE method with deep learning framework was based on the deep neural network (DNN) with log-power spectra features (LPS) as inputs (noisy LPS) and targets (clean LPS) [5].

To improve the basic DNN-based method, numerous methods were proposed. There was a study on the replacement of the DNN with another neural network architecture that could model the speech characteristic better, namely the convolutional neural network (CNN) [6], used in capturing the local frequency structure of speech signals and the long short-term memory (LSTM) recurrent neural network (RNN) [7-10] used to adaptively exploit the context information (CI) of speech signals.

The other branch of research was based on the masking method. This method estimated the enhanced spectrum by applying some types of mask to the magnitude spectrum. Recently, the phase-sensitive mask (PSM) [11] was studied to reflect the phase information to the enhanced spectrum estimation. In [12], a complex ideal ratio mask (cIRM) was proposed for reconstructing the short time Fourier transform coefficient (STFT) rather than reconstructing magnitude spectrum only [12].

In our previous work [13], we proposed a two-stage network (TSN) to utilize the boosting strategy (BS), within a single model. This BS was one of ensemble methods that obtained the final predictions by aggregating the initially obtained multiple base predictions (MBPs) from multiple base predictors. In general, adopting the BS required a high computation cost, as it required MBPs from multiple base predictors [14]. We solved that problem by adopting a single TSN, including prior neural network (pri-NN) to obtain MBPs, and posterior neural network (post-NN) to aggregate them for final prediction [13]. The details of the TSN framework will be discussed in the next section.

Though TSN showed a promising performance in its perceptual quality, we used the DNN for pri-NN, which could utilize limited CI from a fixed size context window. To boost the ability to use the CI for pri-NN, LSTM-RNN can be a good design choice for pri-NN. In addition, TSN modified the noisy magnitude spectrum in the course of enhancement and used the noisy phase for waveform reconstruction, which could cause the spectrogram inconsistency problem and degrade the perceptual quality of enhanced speech signals [15-16].

In this paper, we extended our previous research and the main differences with our previous work are: i) We adopted LSTM-RNN for pri-NN to use CI efficiently. ii) When we reconstructed the waveform from enhanced magnitude spectra, we also reconstructed the phase information to alleviate the spectrogram inconsistency problem rather than simply using the noisy phase. iii) As the speech recognition task is part of key applications for SE, we further investigated the phone recognition performance of proposed methods as well as the perceptual quality of enhanced speech signals.

## 2. Proposed Speech Enhancement System

The objective of rTSN framework is to predict the enhanced LPS from noisy ones based on the BS. For the BS, rTSN has two stages, the pri-NN and post-NN. At the first stage, pri-NN estimates MBPs. Then, the post-NN aggregates MBPs to obtain better final predictions. The multi-objective learning method (MOL) in rTSN framework enables it use the BS by assigning the prediction ability to the pri-NN. When we reconstructed the waveform from the enhanced LPS, we iteratively estimated the phase related to the enhanced LPS based on Griffin-Lim algorithm (GLA) rather than just using the noisy phase [13].

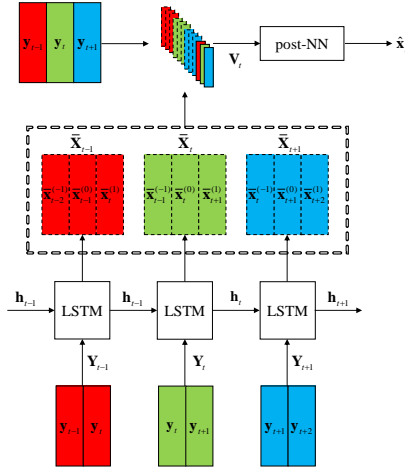

Figure 1: *Description of rTSN when τ is set to 1. Red, green, and blue rectangles with solid borders correspond to $Y_{t-1}$, $Y_t$, and $Y_{t+1}$, respectively, the inputs of pri-NN. Those with the dashed borders correspond to $\bar{X}_{t-1}$, $\bar{X}_t$, and $\bar{X}_{t+1}$, respectively, the outputs from the pri-NN. $V_t$ containing 12 LPS vectors is the input into the post-NN. The post-NN output $\hat{x}_t$ is the estimated $x_t$ corresponding to the clean LPS.*

## 2.1. Recurrent two stage network

### 2.1.1. Prior neural network

As described in Figure 1, the pri-NN initially estimates multiple enhanced LPS $\bar{X}_t$ from noisy ones $Y_t$ using the LSTM-RNN as follows:

$$Y_t = [y_t, y_{t+1}, ..., y_{t+\tau}] \qquad (1)$$

$$\bar{X}_t, h_t = LSTM(Y_t, h_{t-1}) \qquad (2)$$

$$\bar{X}_t = \left[\bar{x}_{t-\tau}^{(-\tau)}, ..., \bar{x}_t^{(0)}, ..., \bar{x}_{t+\tau}^{(-\tau)}\right] \qquad (3)$$

where, $y_t \in \mathbb{R}^N$ is the $t$-th frame noisy LPS vector, N is the feature dimension, $\tau$ is the half window size, $\bar{x}_{t+m}^{(m)} \in \mathbb{R}^N$ is the enhanced LPS vector, and $h_t$ is the hidden activation of the LSTM cell. The main differences between the pri-NN and the ordinary LSTM-RNN are that: i) The pri-NN uses $Y_t$ as its inputs, i.e., future CI, instead of $y_t$ as pri-NN should predict $\bar{X}_t$ including τ future frames at $t$-th step. As LSTM-RNN can naturally model past CI, we do not splice past frames for pri-NN's inputs. Although adopting bi-directional LSTM-RNN can be another option for predicting τ future frames at $t$-th step, however, that kind of network architecture results in high latency, restricting some online applications (e.g. online speech recognition). ii) The pri-NN produces multiple enhanced LPS $\bar{X}_t$ instead of a single enhanced LPS vector, as we can get MBPs for $t$-th frame by aggregating neighboring frames' predictions as follows:

$$\left\{\bar{x}_t^{(m)}\right\}_{m=-\tau}^{\tau} \subset \left\{\bar{X}_{t+m}\right\}_{m=-\tau}^{\tau} \qquad (4)$$

where, $\left\{\bar{x}_t^{(m)}\right\}_{m=-\tau}^{\tau}$ is MBPs for $t$-th frame.

The motivations for adopting LSTM-RNN for the pri-NN instead of the DNN are: i) LSTM-RNN can utilize CI more flexibly than DNN via a fixed sized window, as LSTM has memory cell structures that adaptively control and manage the usage of some of CI. ii) The pri-NN's neighboring outputs are highly overlapped, e.g., $\bar{X}_t$ and $\bar{X}_{t+1}$ include predictions for $t-\tau$ to $t+\tau$ frames and for $t-\tau+1$ to $t+\tau+1$ frames, respectively, so that $2\tau$ frames are overlapping in this case, which implies that neighboring outputs are highly correlated. Therefore, we assume that it assists propagation of some information used for previous predictions for the next step of predictions through recurrent connection of LSTM-RNN.

### 2.1.2. Posterior neural network

The post-NN aggregates MBPs from pri-NN with some additional LPS vectors and obtains the final prediction as follows:

$$V_t = \left\{\bar{X}_{t+m}, y_{t+m}\right\}_{m=-\tau}^{\tau} \qquad (5)$$

$$\hat{x}_t = post\text{-}NN(V_t) \qquad (6)$$

where, post-NN consists of 1-D convolutional layers, where the convolution is conducted across the frequency dimension. As in (4), $V_t$ includes MBPs $\left\{\bar{x}_t^{(m)}\right\}_{m=-\tau}^{\tau}$ for $t$-th frame. The other enhanced frames with the exception of MBPs are corresponding to their respective neighborhoods with $t$-th frame to enable high correlation with $t$-th frame hence we attached them to $V_t$. Additionally, by adding $\left\{y_{t+m}\right\}_{m=-\tau}^{\tau}$ to $V_t$, post-NN could originally extract some local features, that was unavailable to pri-NN, by the convolution layer's local filtering property based on its local connectivity, the different architecture used in pri-NN, which had full connections between LSTM cells. Note that newly re-extracting features with different architecture from noisy LPS is reasonable as enhanced frames from pri-NN could lose some information related with clean LPS. We found that the feature joining method (5) for post-NN improved the performance (in perceptual quality) compared to using MBPs only [13].

### 2.1.3. Multi-Objective learning method (MOL)

The MOL assigns the prediction ability to post-NN as well as pri-NN by formulating the loss function as follows:

$$J_{loss} = J_{post} + \lambda J_{pri} = \sum_t \|\hat{x}_t - x_t\|^2 + \lambda \|\bar{X}_t - X_t\|_F^2, \qquad (7)$$

$$X_t = [x_{t-\tau}, \cdots, x_t, \cdots, x_{t+\tau}], \qquad (8)$$

where $x_t \in \mathbb{R}^N$ is the clean LPS vector, $\lambda$ is a weight for $J_{pri}$, $\|\cdot\|$, and $\|\cdot\|_F$ are $l_2$ and the Frobenius norm, respectively. Note that if the last term is discarded, i.e., $\lambda$ is set to 0, the pri-NN loses its prediction ability, i.e., BS effect is removed, thereby degrading the performance (in perceptual quality) [13].

## 2.2. Phase reconstruction with Griffin-Lim algorithm

From now on, we assume that the LPS to magnitude conversion operation was applied to $\hat{x}_t$ and $y_t$ but we will omit the additional notation for that operation, for simplicity.

In order to reconstruct the time-domain signal from a magnitude spectrogram $\hat{X}$ consisting of $\hat{x}_t$, corresponding phase information $\angle\hat{X}$ is necessary. However, we do not know $\angle\hat{X}$, hence, the corresponding noisy phase information $\angle Y$ is generally used, which causes the inconsistency spectrogram problem as follows:

**Algorithm I** – *Griffin-Lim algorithm*

1: **Initialization**: $X^{[0]} = \hat{X} \cdot e^{j\angle Y}$
2: **for** $i = 1 : K$
3:     $x^{[i]} = ISTFT\left(X^{[i-1]}\right)$
4:     **If** $i = K$, **return** $x^{[i]}$
5:     **else**
6:        $X^{[i]} = \hat{\mathbf{x}} \cdot STFT\left(x^{[i]}\right) / \left|STFT\left(x^{[i]}\right)\right|$
7:     **end**
8: **end**

$$STFT\left(ISTFT\left(\hat{X}e^{j\angle Y}\right)\right) \neq \hat{X}e^{j\angle Y}, \quad (9)$$

where STFT and ISTFT are short-time Fourier transform and its inverse, respectively. The inconsistency spectrogram problem (9) is because $\hat{X}e^{j\angle Y}$ is not a result of applying STFT to some time-domain signal, i.e., $\hat{X}e^{j\angle Y}$ is not true STFT, therefore applying STFT to some time-domain signal (left-hand side in (9)) could not be same with $\hat{X}e^{j\angle Y}$. To estimate $\angle \hat{X}$, we adopt GLA when given $\hat{X}$. However, the major problems of GLA reported [16] are (i) GLA does not include an algorithm for estimating good initial point to start the iteration. (ii) The convergence requires many iterations (> 50, in general), which causes the computation cost problem. To mitigate these problems, we exploit $\angle Y$ as an initial estimation rather than an arbitrary phase. Although $\angle Y$ has distorted phase information from noise, it can locally have phase information related to the clean signal, which can be highly correlated with $\angle \hat{X}$. The used GLA is described in Algorithm I, where, '·' and '/', are element-wise multiplication and division respectively, and '| · |' is the absolute operation.

We experimentally found that using $\angle Y$ as an initial estimation drastically decreased the number of iterations for the convergence of GLA.

## 3. Experiments

### 3.1. Experimental setup

#### 3.1.1. Dataset preparation

For the training speech dataset, 4620 utterances from TIMIT [17] training set were used. For training noise dataset, 150 noise types were used; 100 noise types were recorded by HU [18] and 50 noise types were randomly selected from a sound effect library [19]. One utterance and noise type were picked randomly from training speech and noise dataset, respectively, to build the noisy speech signal. When adding the noise to the speech signal, the FaNT tool [20] was used and SNR (Signal-to-noise ratio) was randomly chosen between -5 and 20 dB. We repeated this procedure until our whole training dataset was approximately 50 h long; 90% of the training dataset was used for training and the remaining 10 % for validation.

For the test speech dataset, we used 192 utterances from TIMIT core test set. For test noise dataset, we used all 15 types of noises from NOISE-92 corpus [21]. The SNRs for test dataset were set to -5, 0, 5, and 10 dB. In addition, we investigated the case when clean speech signal is used.

All aforementioned speech and noise signals were down sampled to 8 kHz. The frame length and shift were set to 25 ms and 10 ms, respectively. The FFT point for LPS was set to 256 so that N was 129. The input LPS were z-score-normalized, as was the output LPS using mean and variance from the input LPS. For methods not using GLA, the noisy phase information was directly used when reconstructing the enhanced LPS to the signal. For GLA, K was set to 5.

#### 3.1.2. Baseline method

For baseline methods, DNN [5] and TSN [13] were adopted. Note that our previously proposed TSN outperformed recently proposed state-of-the-art methods such as fully convolutional neural network [22] and LSTM-RNN [7] based SE systems. The DNN consisted of 3 hidden layers with 2048 hidden units. The past, 4 future frames, and a current frame were spliced the input features of the DNN. The scaled exponential linear units (SELUs) [23] was used for activation function of the DNN. For TSN, we followed model specification proposed in past research [13]. The approximate number of parameters (in millions) was 11.03 and 4.13 for DNN and TSN, respectively.

#### 3.1.3. Recurrent two stage network setup

The pri-NN in rTSN was the uni-directional LSTM-RNN, consisting of 2 hidden layers and each layer had 512 LSTM cells. The post-NN in rTSN was a fully convolutional neural network, which was built with four convolution layers. The kernel size was set to 5 for all convolution layers. The number of output feature maps for convolution layers in post-NN were 256, 128, 64, and 1. SELUs was used for the post-NN's activation function. The approximate number of parameters (in millions) was 5.12 for rTSN. $\tau$ and $\lambda$ were set to 4 and 10, respectively. Note that we found that increasing $\tau$ and $\lambda$ could improve the performance according to our previous research, however, we did not optimize those parameters as our concern was on the effect of proposed architecture and phase reconstruction, therefore we used the values for $\tau$ and $\lambda$, the same with the TSN. The other aforementioned setup was found from our validation dataset.

#### 3.1.4. Training method

The rTSN, TSN, and DNN were trained with Adam [24] using the initial learning rate set to 0.0001. The batch size for TSN and DNN was set to 256. The LSTM in rTSN was unrolled for 64 time-steps and 16 utterances were simultaneously used for training. The early stopping method [25] was applied to decide the number of epochs.

#### 3.1.5. Phone recognizer

In order to verify our proposed SE systems' effect for the noise robust speech recognition task, we conducted the phone recognition experiments. Our phone recognizer was based on DNN with hidden Markov model (DNN-HMM) proposed in a study [26]. The input features of the DNN used 5 left and right context frames including a current one (11 frames in total). For each frame, 13 dimensional Mel-frequency cepstral coefficients (MFCCs) were extracted so that inputs features had 13×11=143 dimensions, which was reduced to 40 dimensions by applying linear discriminant analysis (LDA). Then, z-score normalization was applied to the input features. The number of output units for the DNN was 8859, corresponding to the number of tied tri-phone states. The number of hidden layers was 7 and each layer had 2048 units. For training, the restricted Boltzmann machine-based pre-training was initially conducted. Then, cross entropy loss based supervised training was carried out with alignment information from Gaussian mixture model with HMM (GMM-HMM) trained on clean speech signal. Finally, the DNN was re-trained based on sequence discriminative training referred to as state-level minimum Bayes risk (sMBR) criterion, applied with 5 iterations. The more details can be found in the study [26].

*3.1.6. Evaluation metrics*

To validate the quality and intelligibility of enhanced speech, the perceptual evaluation of speech quality (PESQ) [27] and short-time objective intelligibility (STOI, in %) [28] were used. To evaluate the effect of enhanced speech to the phone recognition task, phone error rate (PER, in %) [29] was used.

**3.2. Experimental results and discussion**

Table 1 summarized our experimental results. Both rTSN and rTSN with GLA outperformed all baseline methods with respect to PESQ and STOI in all SNRs, implying that adopting LSTM-RNN as pri-NN was effective compared to the DNN. This result is elementary, as we found that obtaining better prior base predictions is the most important part within the TSN framework [13] and LSTM-RNN is known to outperform the DNN by utilizing CI more efficiently [8].

Applying GLA for TSN and rTSN shows the consistent performance improvement in both PESQ and STOI with the exception of the clean case, implying that reconstructing phase information corresponding to enhanced magnitude spectrum was effective. In Figure 2, we compared spectrograms of the enhanced speech signals from TSN and TSN with GLA. As in Figure 2, TSN with GLA suppressed the residual noise in TSN. Note that using noisy phase information leaves some impulse-like noise in the time-domain, even clean magnitude is employed so that reconstructing phase information is necessary for the perceptual quality [16].

To verify PER, our phone recognizer was trained on input features extracted from i) noisy speech signals only corresponding to 'Noisy' in Table 1 and ii) the noisy speech signals and corresponding enhanced signals from each SE method. All SE based phone recognizer outperformed the one trained only with noisy speech signals. In contrast to PESQ and STOI, rTSN and rTSN with GLA could not outperform the TSN in SNRs (5 and 10 dB) and the clean signal. In addition, TSN presented the best result with respect to PER on average. This was because the objective of our SE systems was minimizing MSE between clean and noisy LPS, which was not directly related to the PER, hence there is no guarantee that well enhanced speech signal was directly helpful for the following phone recognizer. In particular, in SE perspective, removing some noisy spectra can be preferred in MSE aspect; however, retaining that spectra can be helpful for the phone recognition if that spectra has some phone-related information.

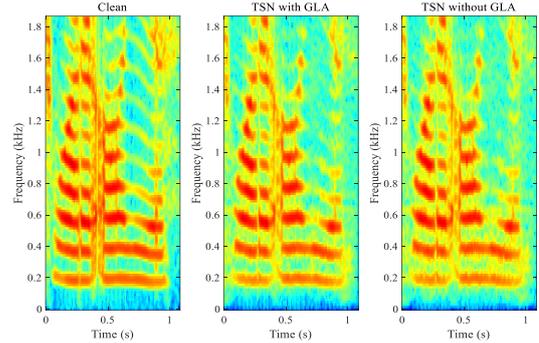

Figure 2: *Example spectrograms for clean (left), enhanced speech signals from TSN with GLA (middle) and TSN without GLA (right). The noise signal was a military vehicle and SNR was -5dB.*

Furthermore, TSN with GLA underperformed compared to TSN in PER, while rTSN with GLA outperformed rTSN, implying that reconstructing phase information is less relevant with improvement of PER although it is consistently effective in PESQ and STOI. This fact is because our phone recognizer uses MFCCs as input features, discarding the phase information in the course of MFCC extraction. Thus, reconstructing the phase information can be redundant if the following speech recognition system does not use the phase-related features.

## 4. Conclusions

We extended TSN framework by adopting LSTM-RNN instead of DNN as pri-NN. Furthermore, the phase information was reconstructed by applying the GLA initialized by noisy phase. In PESQ and STOI, rTSN with GLA outperformed all other baseline methods, however, this improvement could not lead to PER. Therefore, our future work will be to improve the relationship between SE and speech recognition task by adopting multi-task learning method.

Table 1: *Performance comparison. PESQ, STOI and PER are averaged over 15 noise types. The numbers in bold and italic indicate the best and second-best results, respectively.*

| | PESQ | | | | | | STOI | | | | | | PER | | | | | |
|---|---|---|---|---|---|---|---|---|---|---|---|---|---|---|---|---|---|---|
| **SNR** | -5 dB | 0 dB | 5 dB | 10 dB | Clean | Avg. | -5 dB | 0 dB | 5 dB | 10 dB | Clean | Avg. | -5 dB | 0 dB | 5 dB | 10 dB | Clean | Avg. |
| **Noisy** | 1.323 | 1.692 | 2.016 | 2.346 | 4.500 | · | 52.12 | 66.87 | 78.46 | 86.46 | 100.0 | · | 67.73 | 67.83 | 57.84 | 44.75 | **24.34** | 52.50 |
| **DNN** | 1.817 | 2.278 | 2.604 | 2.904 | 3.668 | 2.654 | 55.99 | 71.75 | 80.97 | 86.95 | 93.82 | 77.90 | 53.49 | 53.28 | 45.06 | 38.28 | 30.33 | 44.09 |
| **TSN** | 1.898 | 2.356 | 2.694 | 2.998 | 3.714 | 2.732 | 59.06 | 74.87 | 83.91 | 90.06 | 97.78 | 81.14 | *50.30* | *50.38* | **39.32** | **32.32** | *25.18* | **39.50** |
| **TSN+GLA** | 1.929 | 2.391 | 2.725 | 3.020 | 3.617 | 2.736 | 59.93 | 75.78 | 84.61 | 90.42 | 97.51 | 81.65 | 50.96 | 51.06 | 41.84 | *35.19* | 27.87 | 41.38 |
| **rTSN** | *1.948* | *2.444* | *2.805* | *3.094* | **3.891** | *2.836* | *61.55* | *77.09* | *85.85* | *91.52* | **98.93** | *82.99* | 50.34 | **50.30** | 42.01 | 36.19 | 29.01 | 41.57 |
| **rTSN+GLA** | **1.985** | **2.485** | **2.844** | **3.121** | *3.779* | **2.843** | **62.68** | **78.14** | **86.68** | **91.94** | *98.70* | **83.63** | **50.27** | 50.53 | *41.73* | 35.34 | 28.88 | *41.35* |